\documentclass[a4paper,fleqn,usenatbib]{mnras}

\usepackage{newtxtext,newtxmath}

\usepackage[T1]{fontenc}
\usepackage{ae,aecompl}

\usepackage{booktabs,caption,fixltx2e}
\usepackage[flushleft]{threeparttable}

\usepackage{graphicx}	
\usepackage{amsmath}	
\usepackage{amssymb}	

\usepackage{etoolbox}
\makeatletter
\patchcmd\@combinedblfloats{\box\@outputbox}{\unvbox\@outputbox}{}{%
   \errmessage{\noexpand\@combinedblfloats could not be patched}%
}%
 \makeatother
 
\title[Universe opacity and Type Ia supernova dimming]{Universe opacity and Type Ia supernova dimming}

\author[V. Vavry\v{c}uk]{
V\'{a}clav Vavry\v{c}uk,$^{1}$\thanks{E-mail: vv@ig.cas.cz}
\\
$^{1}$Institute of Geophysics, The Czech Academy of Sciences, Bo\v{c}n\'{i} II, Praha 4, 14100, Czech Republic\\
}

\date{Accepted XXX. Received YYY; in original form May 29, 2019}

\pubyear{2019}

\begin{document}
\label{firstpage}
\pagerange{\pageref{firstpage}--\pageref{lastpage}}
\maketitle
\begin{abstract}
In this paper, I revoke a debate about an origin of Type Ia supernova (SN Ia) dimming. I argue that except for a commonly accepted accelerating expansion of the Universe, a conceivable alternative for explaining this observation is universe opacity caused by light extinction by intergalactic dust, even though it is commonly assumed that this effect is negligible. Using data of the Union2.1 SN Ia compilation, I find that the standard $\Lambda$CDM model and the opaque universe model fit the SN Ia measurements at redshifts $z < 1.4$ comparably well. The optimum solution for the opaque universe model is characterized by the B-band intergalactic opacity $\lambda_B = 0.10 \pm 0.03 \, \mathrm{Gpc}^{-1}$ and the Hubble constant $H_0 = 68.0 \pm 2.5 \, \mathrm{km\,s^{-1}\, Mpc^{-1}}$.  The intergalactic opacity is higher than that obtained from independent observations but still within acceptable limits. This result emphasizes that the issue of the accelerating expansion of the Universe as the origin of the SN Ia dimming is not yet definitely resolved. Obviously, the opaque universe model as an alternative to the $\Lambda$CDM model is attractive, because it avoids puzzles and controversies associated with dark energy and the accelerating expansion. 
\end{abstract}


\section{Introduction}
Type Ia supernovae (SNe Ia) dimming is one of the most exciting discoveries in astronomy in the last 20 years. The first results published by \citet{Riess1998} and \citet{Perlmutter1999} were based on measurements of 16 and 42 high-redshift SNe Ia, respectively. However, observations of the unexpected SNe Ia dimming motivated large-scale systematic searches for SNe Ia and resulted in a rapid extension of supernovae compilations  \citep{Sullivan2011, Suzuki2012, Campbell2013, Betoule2014, Jones2018, Scolnic2018_ApJ}. This surprising phenomenon was explained by an accelerating expansion of the Universe, and consequently a concept of the cosmological constant \citep{Einstein1917,Blome_Priester1985,Carroll1992} was revived and reintroduced as dark energy into the cosmological models \citep{Weinberg1989,Riess2000,Sahni_Starobinsky2000,Peebles_Ratra2003}.

Processing of the SNe Ia data is not straightforward. Prior to interpretations, corrections must be applied to transform an observed-frame magnitude to a rest-frame magnitude. This includes a cross-filter $K$-correction and a correction for light extinction due to absorption by dust in the host galaxy and in our Galaxy  \citep{Perlmutter1999, Nugent2002, Riess2004b}. The uncertainties in the extinction corrections are in general much higher than those in the $K$-corrections \citep{Nugent2002}. Moreover, the uncertainties increase due to neglecting redshift-dependent extinction by intergalactic dust which might display different reddening than the interstellar dust \citep{Menard2010b}. Obviously, these uncertainties raise the question, whether the observations of supernovae dimming are not partly or fully a product of intergalactic extinction. 

Intergalactic opacity as a possible origin of dimming of the SNe Ia luminosity was proposed by \citet{Aguirre1999b, Aguirre1999a} and \citet{Aguirre_Haiman2000}. Also \citet{Menard2010b} point out that a reddening-based correction is not sensitive to intergalactic opacity and it might bias the calculated distance modulus and the cosmological parameters describing the accelerating expansion. This problem was also addressed by \citet{Riess2004b} who found that some models of intergalactic dust might produce a similar dimming as observed and interpreted by the accelerating expansion. The authors fitted a theoretical extinction curve to the SNe Ia dimming in the redshift interval $z < 0.5$. They found a satisfactory fit, but a remarkable discrepancy appeared at higher $z$, see \citet[their fig. 3, model A]{Goobar2002} or \citet[their fig. 7, 'high-$z$ gray dust' model]{Riess2004b}. The discrepancy was removed for a redshift-independent proper density of intergalactic dust for $z > 0.5$, however, this contradicts the idea of an increasing proper density of intergalactic dust due to the smaller volume of the Universe in the past. 

In this paper, I revisit this analysis and argue that rejecting the universe opacity as a possible origin of SNe Ia dimming was not fully justified. I show that intergalactic dust might produce similar effects in the SNe Ia dimming as the accelerating expansion. The proposed opacity model is characterized by an increase of the proper dust density with redshift for all $z$ but not only for $z < 0.5$ as suggested by \citet{Goobar2002} and \citet{Riess2004b}. The model fits the SNe Ia data comparably well as the currently used $\Lambda$CDM model does. Advantageously, the proposed model does not need the controversial dark energy concept and the accelerating expansion.

\section{Intergalactic opacity}
The intergalactic opacity $\lambda_V$ (defined as attenuation $A_V$ of intergalactic space per unit ray path) is caused by light extinction by intergalactic dust and it is spatially and redshift dependent. It is mostly appreciable at close distance from galaxies and in intracluster space. \citet{Menard2010a} report visual intergalactic attenuation $A_V = (1.3 \pm 0.1) \times 10^{-2}$ mag at distance from a galaxy of up to 170 kpc and $A_V = (1.3 \pm 0.3) \times 10^{-3}$ mag at distance of up to 1.7 Mpc. Similar values are observed for the visual attenuation of intracluster dust \citep{Muller2008, Chelouche2007}. An averaged value of intergalactic extinction was measured by \citet{Menard2010a} by correlating the brightness of $\approx$85.000 quasars at $z > 1$ with the position of 24 million galaxies at $z \approx 0.3$ derived from the Sloan Digital Sky Survey. The authors estimated $A_V$ to about 0.03 mag at $z = 0.5$. A consistent opacity was reported by \citet{Xie2015} who studied the luminosity and redshifts of the quasar continuum of $\approx$90.000 objects and estimated the intergalactic opacity at $z < 1.5$ as $\lambda_V \approx 0.02 \,h \, \mathrm{Gpc}^{-1}$.

Extinction by dust can also be measured from the hydrogen column densities of damped Lyman $\alpha$ absorbers (DLAs). Based on the Copernicus data, \citet{Bohlin1978} report a linear relationship between the total hydrogen column density, $N_\mathrm{H} = 2 N_\mathrm{H2} + N_\mathrm{HI}$, and the color excess, $N_\mathrm{H} / E\left(B-V\right) = 5.8 \times 10^{21} \, \mathrm{cm}^{-2} \, \mathrm{mag}^{-1}$, hence
$N_\mathrm{H} / A_V \approx 1.87 \times 10^{21} \, \mathrm{cm}^{-2} \, \mathrm{mag}^{-1}$ for  $R_V = A_V/E\left(B-V\right)= 3.1$, which is a typical value for our Galaxy \citep{Cardelli1989,Mathis1990}. The result has been confirmed by \citet{Rachford2002} using FUSE data, who refined the slope between $ N_\mathrm{H}$ and $E\left(B-V\right)$ to $5.6 \times 10^{21} \, \mathrm{cm}^{-2} \, \mathrm{mag}^{-1}$. Taking into account observations of the mean cross-section density of DLAs, $\langle n \sigma \rangle =  \left(1.13 \pm 0.15 \right) \times 10^{-5} \, h \, \mathrm{Mpc}^{-1}$ \citep{Zwaan2005}, the characteristic column density of DLAs, $N_\mathrm{HI} \approx 10^{21} \, \mathrm{cm}^{-2}$, and the mean molecular hydrogen fraction in DLAs of about $0.4 - 0.6$  (\citet[their Table 8]{Rachford2002}), the intergalactic opacity $\lambda_V$ at $z = 0$ is $\lambda_V \approx 1-2 \times 10^{-5} \, \mathrm{Mpc}^{-1}$, which is the result of \citet{Xie2015}.

A low value of intergalactic opacity $\lambda_V \approx 0.02 \, \mathrm{Gpc}^{-1}$ indicates that the intergalactic space is almost transparent at $z = 0$. However, intergalactic opacity is redshift dependent. It increases with redshift and a transparent universe becomes significantly opaque at high redshifts. The opacity at high redshifts is caused by a high proper dust density due to the small volume of the Universe in the past. Since the dust density increases with redshift as $(1+z)^3$, the opacity increase is enormous and the total attenuation $A_V$ can achieve a value of $0.2$ mag at $z = 1$ or even a value of $0.7-0.8$ mag at $z = 3$  \citep[his fig. 10a]{Vavrycuk2018}.

\begin{figure*}
\includegraphics[angle=0,width=10 cm,trim=100 40 100 50,clip = true]{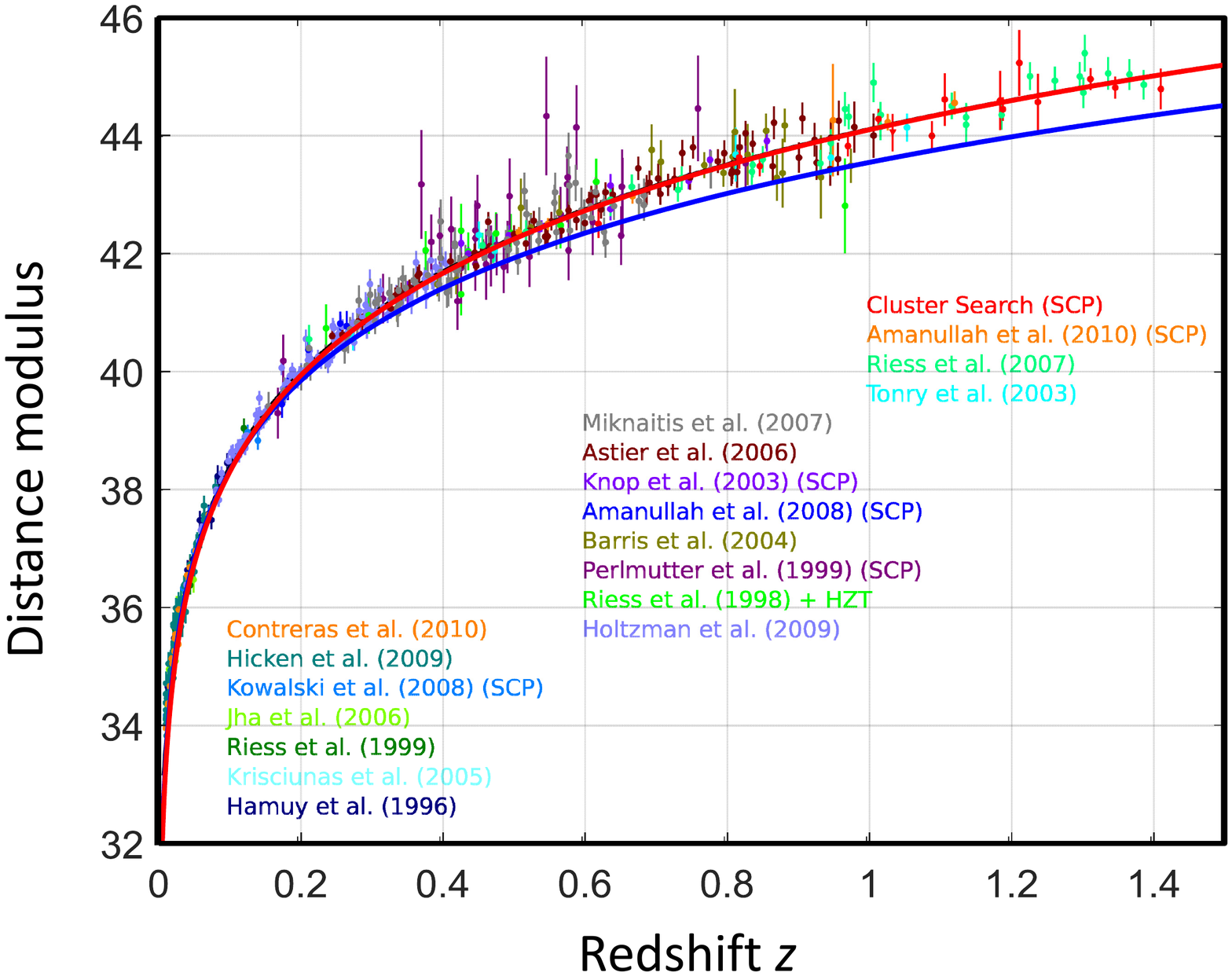}
\caption{
The Hubble diagram with SNe Ia measurements (Union2.1 dataset). Solid red line - the $\Lambda$CDM model with $\Omega_m = 0.3$, $\Omega_\Lambda = 0.7$, and $\Omega_k = 0$. Solid blue line - the cosmological model with $\Omega_m = 1.0$, $\Omega_\Lambda = 0$, and $\Omega_k = 0$. The data are not corrected for intergalactic opacity. Modified after \citet[their fig. 4]{Suzuki2012}.
}
\label{fig:1}
\end{figure*}

\section{Fitting supernovae measurements}

The current supernovae compilations are comprised of about one thousand SNe Ia discovered and spectroscopically confirmed \citep{Sullivan2011, Suzuki2012, Campbell2013, Betoule2014, Rest2014, Scolnic2018_ApJ,Riess2018}. Every SN Ia is described by its apparent rest-frame B-band magnitude, light curve shape, and colour correction. These parameters are used in the Tripp formula \citep{Tripp1998,Guy2007} for determining the distance modulus $\mu(z)$ (Fig.~\ref{fig:1}),
\begin{equation}\label{eq1}
\mu = m_B - M_B + \alpha x_1 - \beta c \,,
\end{equation}
where $m_B$ is the apparent rest-frame B-band magnitude, $M_B$ is the absolute B-band magnitude, $c$ and $x_1$ are the colour and stretch parameters, respectively, and the coefficients $\alpha$ and $\beta$ are the global nuisance parameters to be calculated when seeking an optimum cosmological model. The distance modulus $\mu$ is related to the expansion history by the following equation,
\begin{equation}\label{eq2}
\mu = 25 + 5 \mathrm{log}_{10}\left(d_L\right) \,,
\end{equation}
where $d_L$ is the luminosity distance (in Mpc), which is expressed in flat ($\Omega_k = 0$) space as (e.g., \citeauthor{Subramani2019} \citeyear{Subramani2019}, their equation 12) 
\begin{equation}\label{eq3}
d_L = \left(1+z\right)\int^{z}_0\frac{c dz'}{H\left(z'\right)} \,.
\end{equation}
%
\begin{figure*}
\includegraphics[angle=0,width=15 cm,trim=50 160 100 120, clip=true]{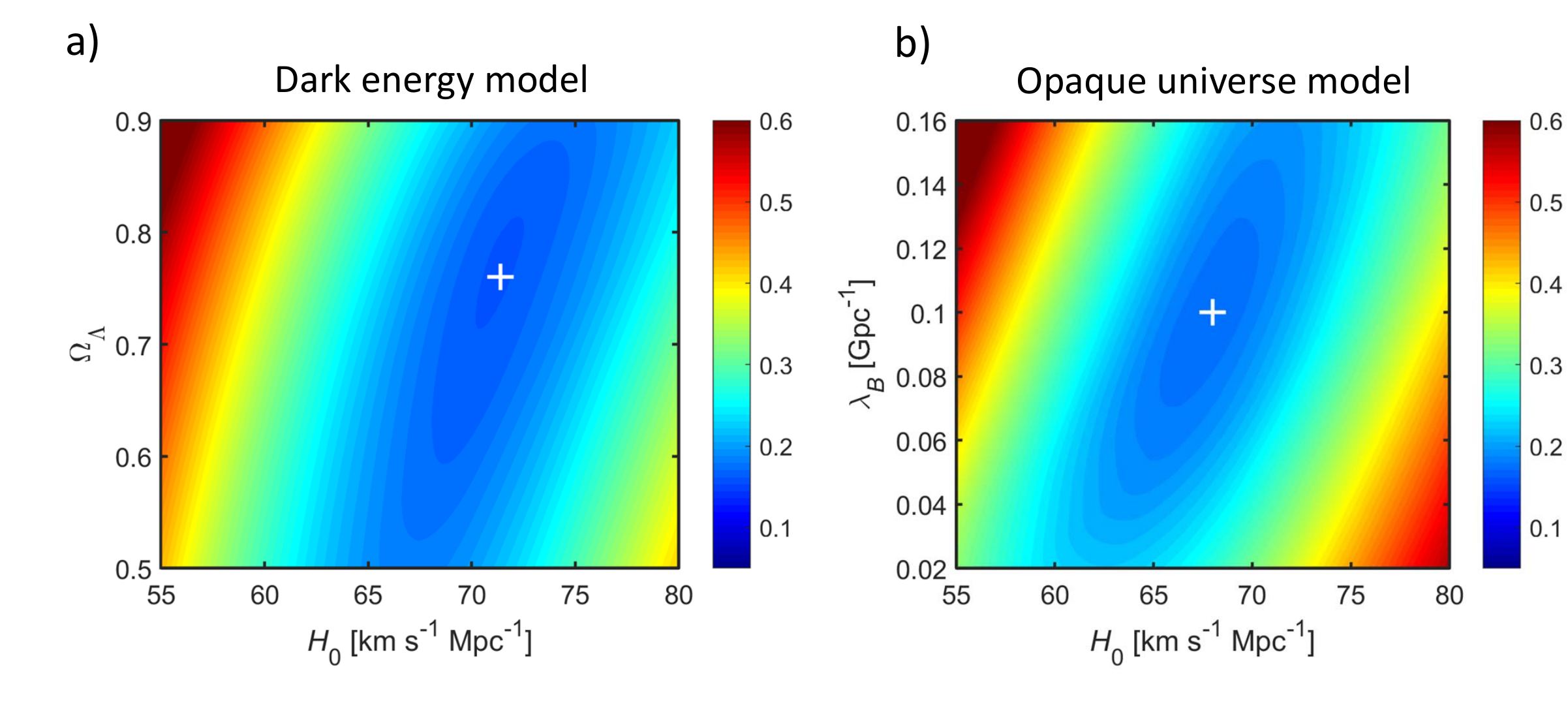}
\caption{
Inversion for optimum cosmological parameters of (a) the $\Lambda$CDM model and (b) the opaque universe model. The colour shows the mean of the absolute distance modulus residua between the predicted model and  the SNe Ia data as a function of: (a) the Hubble constant $H_0$ and the dark energy $\Omega_\Lambda$, and (b) the Hubble constant $H_0$ and the B-band intergalactic opacity $\lambda_B$. Only the most accurate 552 SNe Ia with $z$ < 1.4 and an error less than 0.50 mag are used. The optimum solutions marked by the white plus signs are defined by: (a) $H_0 = 71.4 \,\, \mathrm{km \, s^{-1} \, Mpc^{-1}}$ and $\Omega_\Lambda = 0.77$ ($\alpha = 0.122$ and $\beta = 2.466$), and (b) $H_0 = 68.0 \,\, \mathrm{km \, s^{-1} \, Mpc^{-1}}$ and $\lambda_B = 0.10 \,\, \mathrm{Gpc^{-1}}$ ($\alpha = 0.088$ and $\beta = 1.920$).
}
\label{fig:2}
\end{figure*}

\subsection{$\Lambda$CDM model}
The standard $\Lambda$CDM cosmological model \citep{Planck_2015_XIII} for the matter-dominated universe is described by the following equation
\begin{equation}\label{eq4}
H^2\left(a\right) = H^2_0 \left[{\Omega_m a^{-3} + \Omega_\Lambda + \Omega_k a^{-2}}\right] \,,
\end{equation}
where
\begin{equation}\label{eq5}
\Omega_m + \Omega_\Lambda + \Omega_k = 1 \,.
\end{equation}
The function $H(a)$ is the Hubble parameter characterizing the universe expansion with the scale factor $a = 1/(1+z)$, $\Omega_m$ is the total matter density contribution, $\Omega_k$ is related to the curvature of the Universe, and $\Omega_\Lambda$ is the dark energy contribution. Since measurements indicate that the Universe is nearly flat,  $\Omega_k$ is zero in equation (4). Imposing the second time derivative of the scale factor $a$ to be zero, $\ddot{a} = 0$, the transition from decelerating to accelerating expansion occurs at 
\begin{equation}\label{eq6}
a=\left(\frac{\Omega_m}{2\Omega_\Lambda}\right)^{1/3} \,,
\end{equation}
which  yields values $a = 0.69$ and $z = 0.67$ for the commonly used parameters  $\Omega_m = 0.3$ and $\Omega_\Lambda = 0.7$.
%

\subsection{Opaque universe model}
The universe opacity is quantified by the redshift-dependent optical depth, which is expressed in the B-band as follows \citep[his equation 19]{Vavrycuk2017a}
\begin{equation}\label{eq7}
\tau_B = \int^{z}_0\lambda_B \left(1+z'\right)^2 \frac{c dz'}{H\left(z'\right)} \,.
\end{equation}
where the Hubble parameter $H(z)$ is defined in equation (4) with $\Omega_m = 1.0$, $\Omega_\Lambda = 0$ and $\Omega_k = 0$. The parameter $\lambda_B$ is the rest-frame B-band attenuation per unit ray path. Equation (7) takes into account an increase of the proper dust density with redshift as $(1+z)^3$ and a decrease of frequency-dependent opacity with $z$ as $(1+z)^{-1}$ due to the $1/\lambda$ extinction law \citep{Mathis1990}.  Since wavelengths at the observer are longer due to redshift, they are less attenuated at the observer than at the source. In addition, a proper distance decreases with $z$ as $(1+z)^{-1}$, but this is eliminated by an increase of light extinction with $z$ as $(1+z)$ because the arrival rate of the photons suffers time dilation, often called the energy effect \citep{Roos2003}. 

Finally, the extinction correction of the distance modulus is expressed as 
\begin{equation}\label{eq8}
\Delta \mu = -2.5 \, \mathrm{log}_{10} \, e^{\tau_B(z)} \,.
\end{equation}
%

\subsection{Optimum model parameters}

For fitting the two mentioned cosmological models, the Union2.1 SNe compilation \citep{Suzuki2012} is used. The free parameters of the $\Lambda$CDM model (dark energy model) are the Hubble constant $H_0$ and the dark energy $\Omega_\Lambda$. The free parameters of the opaque universe model are the Hubble constant $H_0$ and the B-band opacity $\lambda_B$ ($\Omega_\Lambda = 0$). The optimum values of these parameters are found by a grid search. Only 552 most accurate SNe Ia with a distance modulus error less than 0.5 mag are considered. The mean distance modulus error is 0.2 mag. In order for the residua to have the same weight for different redshift intervals with a different number of the SNe Ia measurements, the mean absolute values of the residua are calculated in the following redshift bins: $z$ = [0, 0.025, 0.05, 0.1, 0.15, 0.20, 0.25, 0.30, 0.35, 0.40, 0.45, 0.5, 0.6, 0.8, 1.0, 1.2, 1.4]. The width of bins is not uniform being narrower at low redshifts. In this way, the distribution of the SNe Ia in bins is more uniform having, respectively, the following numbers: [65, 75, 35, 27, 28, 27, 38, 33, 27, 28, 20, 37, 48, 37, 16, 11]. 

The misfit functions have a very shallow minimum for both considered models. For the opaque universe model, the best fit is found in the redshift interval 0 < $z$ < 1.4 for $H_0 = 68.0 \pm 2.5 \,\, \mathrm{km \, s^{-1} \, Mpc^{-1}}$ and $\lambda_B = 0.10 \pm 0.03 \,\, \mathrm{Gpc^{-1}}$ (Fig.~\ref{fig:2}b). Analogously, the same data inverted for the optimum values of $H_0$ and $\Omega_\Lambda$, describing the  $\Lambda$CDM model, yield the best fit for $H_0 = 71.4 \pm 2.5 \,\, \mathrm{km \, s^{-1} \, Mpc^{-1}}$ and $\Omega_\Lambda = 0.77 \pm 0.13$ (Fig.~\ref{fig:2}a). The uncertainty limits were calculated from the misfit function when all solutions with a fit at least of 99\% of the best fit were accepted. 

\begin{figure*}
\includegraphics[angle=0,width=13cm,trim=50 20 80 40, clip = true]{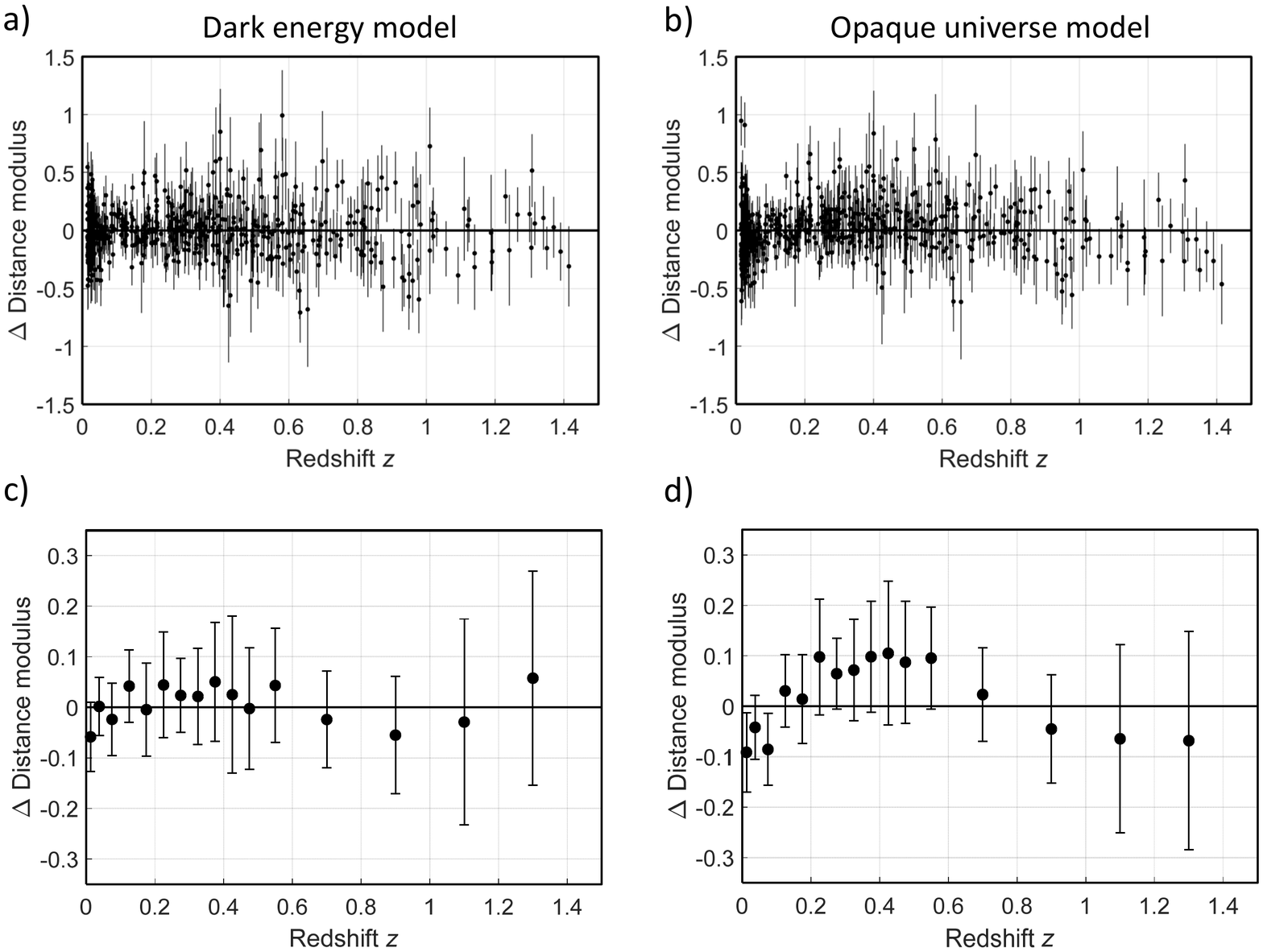}
\caption{
Residual Hubble plots for (a-b) the individual SNe Ia data and (c-d) the binned SNe Ia data. (a,c) The flat $\Lambda$CDM model with $H_0 = 71.4 \,\, \mathrm{km \, s^{-1} \, Mpc^{-1}}$ and $\Omega_\Lambda = 0.77$.  (b,d) The opaque universe model with $\lambda_B = 0.10 \,\, \mathrm{Gpc^{-1}}$ and $H_0 = 68.0 \,\, \mathrm{km \, s^{-1} \, Mpc^{-1}}$. The error bars in (c-d) show the 95\% confidence intervals. Data are taken from \citet{Suzuki2012}.
}
\label{fig:3}
\end{figure*}

The residual Hubble plots are shown in Fig.~\ref{fig:3}. The optimum solutions have roughly comparable misfits for both models: 0.16 for the $\Lambda$CDM model (Fig.~\ref{fig:3}a), and 0.17 for the model of the opaque universe (Fig.~\ref{fig:3}b). However, as illustrated in Fig.~\ref{fig:3}c,d, the $\Lambda$CDM model performs slightly better than the opaque universe model, because of a better fit of the binned residua for redshifts less than 0.6. Nevertheless, the model of the opaque universe also accounts for the observed dimming of the SNe Ia within the confidence level of 95\% (Fig.~\ref{fig:3}d).

\begin{figure*}
\includegraphics[angle=0,width=13cm,trim=50 160 80 90,clip = true]{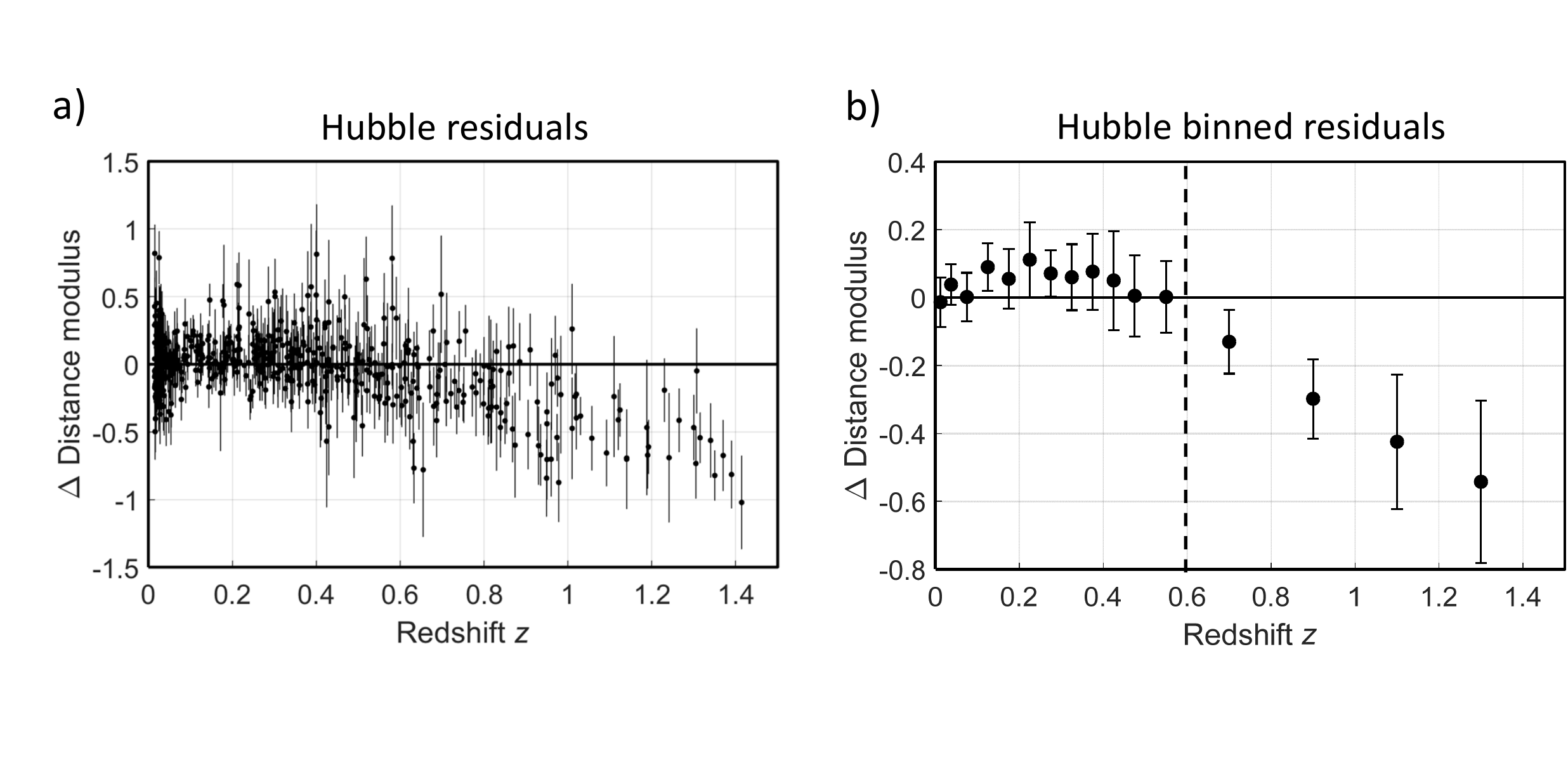}
\caption{
Residual Hubble plots for (a) the individual and (b) binned SNe Ia data for the opaque universe model with $\lambda_B = 0.18 \,\, \mathrm{Gpc^{-1}}$ and $H_0 = 70.4 \,\, \mathrm{km \, s^{-1} \, Mpc^{-1}}$. The vertical dashed line in (b) denotes the upper redshift limit of the SNe Ia data considered in the inversion. The error bars in (b) show the 95\% confidence intervals. Data are taken from \citet{Suzuki2012}.
}
\label{fig:4}
\end{figure*}

Importantly, the optimum solutions are sensitive to the redshift interval of the SNe Ia used in the inversion. If the optimum model of the opaque universe is searched using the SNe Ia with redshifts $z \leq 0.6$ only, and the SNe Ia are grouped in the following redshift bins: $z$ = [0, 0.025, 0.05, 0.1, 0.15, 0.20, 0.25, 0.30, 0.35, 0.40, 0.45, 0.5, 0.6], we get $H_0 = 70.4 \pm 2.0 \,\, \mathrm{km \, s^{-1} \, Mpc^{-1}}$ and $\lambda_B = 0.18 \pm 0.05 \,\, \mathrm{Gpc}^{-1}$ (Fig.~\ref{fig:4}). We see in Fig.~\ref{fig:4}b that the model fits the SNe measurements at $z < 0.6$ much better than in Fig.~\ref{fig:3}d, but a significant discrepancy appears at higher redshifts. A similar result was obtained by \citet{Goobar2002} and \citet{Riess2004b}, who concluded that the discrepancy at high redshifts excludes the SNe Ia dimming to be produced by intergalactic dust.

Note that the best estimate of $H_0$ for the $\Lambda$CDM model obtained by \citet{Riess2016} using the SNe Ia data is $73.24 \pm 1.74 \,\, \mathrm{km \, s^{-1} \, Mpc^{-1}}$. The precision of the distance scale was further improved by a factor of 2.5 by \citet{Riess2018}. This value is within the uncertainty of the estimate of $H_0$ for the $\Lambda$CDM model obtained in this paper and shown in Fig.~\ref{fig:2}a. The accuracy of the presented value is, however, lower, because no specific analysis of $M_B$ (which is degenerate with $H_0$) was performed and no further selection of the most accurate SN Ia data was applied before the inversion. Interestingly, the Hubble constant $H_0$ in the $\Lambda$CDM model obtained from the Planck measurements of the cosmic microwave background is $H_0 = 67.4 \pm 0.5 \,\, \mathrm{km \, s^{-1} \, Mpc^{-1}}$ \citep{Planck_2018_VI}, being inconsistent with the results of  \citet{Riess2018} and pointing to another  difficulty of the $\Lambda$CDM model \citep{Mortsell_Dhawan2018}.

\section{Discussion and conclusions}

The $\Lambda$CDM model explains satisfactorily current observations of the SNe Ia dimming by introducing dark energy into the Friedmann equations and considering the accelerating expansion of the Universe. Both concepts invoke, however, essential difficulties in physics. Dark energy causes negative pressure, which is required to explain the accelerating expansion. The pressure is extremely small, its magnitude being by 120 orders lower than a theoretical value predicted by quantum field theory \citep[his equation 5]{Koyama2016}. It is also unclear, why the dark energy has the same order of magnitude as the matter energy density just at the present epoch, although the matter density changed by a factor of $10^{42}$ during the evolution of the universe \citep{Weinberg2013, Bull2016}. Tension with the flat $\Lambda$CDM model arises also from a high-redshift Hubble diagram of supernovae, quasars, and gamma-ray bursts for redshifts 1.4 < $z$ < 5 \citep{Lusso2019ArXiv}.  In addition, the dark energy concept predicts the speeds of gravitational waves and light to be generally different \citep{Sakstein2017}. However, observations of the binary neutron star merger GW170817 and its electromagnetic counterparts proved that both speeds coincide with a high accuracy ($< 5 \times 10^{-16}$). Hence, most of the dark energy models are disfavored \citep{Ezquiaga2017}. Other observational challenges to the $\Lambda$CDM model are summarized in \citet{Kroupa2012}, \citet{Kroupa2015}, \citet{Buchert2016} and \citet{Bullock2017}.

By contrast, light extinction by intergalactic dust as a physically plausible origin of the SNe Ia dimming has been ignored or underrated. Its possible role  in the SNe Ia dimming was discussed by several authors including \citet{Perlmutter1999} and \citet{Riess1998}, but an opinion prevailed that the  effect of intergalactic dust on the SNe Ia observations is minor. However, recent detailed studies of dust extinction in the SNe Ia spectra revealed that this issue is more complicated than so far assumed and that the standard extinction corrections applied uniformly to the SNe Ia observations were too simplified. For example, unexpected complexities were detected in reddening, being characterized by a variety of extinction laws with $R_V$ going down to 1.4 \citep{Amanullah2014, Amanullah2015, Gao2015}. Consequently, approximate reddening-based corrections might bias the calculated distance modulus and lead to misinterpretations and to erroneously neglecting the role of intergalactic dust in the SNe Ia dimming \citep{Menard2010b}. 

This argument is supported by modelling performed in this paper, which shows that the opaque universe model can fit the observations almost equally well as the $\Lambda$CDM model, if inverted in the whole redshift interval of the SNe Ia observations. The found optimum B-band intergalactic opacity is $\lambda_B = 0.10 \pm 0.03 \,\, \mathrm{Gpc^{-1}}$ and the Hubble constant is $H_0 = 68.0 \pm 2.5 \, \mathrm{km \, s^{-1} \, Mpc^{-1}}$. Since the redshift-distance relation differs in the $\Lambda$CDM model and in the opaque universe model by a factor of about $1.5-2$, the value $\lambda_V \approx 0.02 \, h \, \mathrm{Gpc}^{-1}$ of \citet{Xie2015} recalculated to the opaque universe model is $\lambda_V \approx 0.04 \, \mathrm{Gpc}^{-1}$. Taking into account that $\lambda_B$ is expressed as $\lambda_B = \lambda_V (R_V+1)/R_V$, we get $\lambda_B \approx 0.05 \, \mathrm{Gpc}^{-1}$ for $R_V = 3.1$ \citep{Cardelli1989}. Hence, the retrieved value of $\lambda_B \approx 0.10 \pm 0.03 \,\, \mathrm{Gpc^{-1}}$ is higher than that expected from  independent observations but still within reasonable limits. 

The presented results emphasize that the issue of the accelerating expansion of the Universe is not yet definitely resolved and it should be revisited in a more thorough way in future. Obviously, the opaque universe model as an alternative to the $\Lambda$CDM model is attractive, because it avoids puzzles and controversies associated with dark energy and the accelerating expansion. The opaque universe model can straightforwardly explain anisotropic Hubble residuals by assuming a slightly anisotropic distribution of intergalactic dust instead of interpreting them by anisotropic expansion of the Universe  \citep{Schwarz_Weinhorst2007, Wang_Wang2014, Javanmardi2015}. The model also successfully explains phenomena related to the extragalactic background light, such as its bolometric intensity and the luminosity density evolution with redshift \citep{Vavrycuk2017a}, and properties of the cosmic microwave background \citep{Vavrycuk2018}.
 
\section*{Acknowledgements} 
I thank an anonymous reviewer for his detailed and helpful comments and Nao Suzuki for permitting to reproduce fig. 4 from Suzuki et al. (2012).


\bibliographystyle{mnras}
\bibliography{paper} 

`	\end{document}